\documentclass[a4paper,11pt]{article}

\usepackage[T1]{fontenc}
\usepackage{lmodern}
\usepackage{amsmath,amssymb}
\usepackage{geometry}
\usepackage[colorlinks=true,linkcolor=blue,citecolor=blue,urlcolor=blue]{hyperref}
\usepackage{tikz}
\usetikzlibrary{cd}
\usepackage{slashed}
\usepackage{pgfplots}
\pgfplotsset{compat=1.17} 
\usetikzlibrary{arrows.meta}
\usetikzlibrary{decorations.markings}
\usetikzlibrary{calc}

\usepackage[p,osf]{cochineal}
\usepackage[scale=.95,type1]{cabin}
\usepackage[cochineal,bigdelims,cmintegrals,vvarbb]{newtxmath}
\usepackage[zerostyle=c,scaled=.94]{newtxtt}
\usepackage[cal=boondoxo]{mathalfa}

\newcommand{\dpartial}[2]{\frac{\partial #1}{\partial #2}} 

\begin{document}

\title{\textbf{Scaling Relations, Morphological Stability, and Asymptotic Freedom of Plasma-Surface Deposition Dynamics}}
\author{{\small Joel Saucedo$^{1}$, Uday Lamba$^{2}$, Hasitha Mahabaduge$^{1}$}\\[4pt]
{\small $^{1}$Department of Chemistry, Physics, \& Astronomy, Georgia College \& State University, GA, 31061, USA}\\[-2pt]
{\small $^{2}$ Department of Physics \& Astronomy, Ithaca College, NY, 14850, USA}}
\date{}

\maketitle

\begin{abstract}
Connecting plasma processing parameters to the resultant film microstructure remains a fundamental challenge in materials synthesis, one that has largely confined process design to empirical approaches. To bridge this gap, we develop a predictive analysis of coupling by applying a renormalization group (RG) analysis to an effective Hamiltonian for the stochastic dynamics of the plasma-surface interface, derived systematically from microscopic principles.

The central result from this formalism is the system's exhibition of asymptotic freedom; the effective dimensionless coupling, $g$, between the plasma and the growing surface is found to weaken systematically at macroscopic length scales, a finding that provides a rigorous justification for the success of continuum-level models in describing large-scale film evolution. The RG framework yields a non-perturbative scaling relation for the mean grain area, $\langle A \rangle \propto \exp(\kappa/g)$, where $g$ itself is defined by fundamental parameters such as ion flux ($\Phi$) and ion collision time ($\tau_{\text{ion}}$). This relation reveals the origin of widely-observed empirical power-law scaling, showing it to be an effective behavior limited to specific process regimes. Crucially, the model furnishes sharp, testable predictions, including the pressure-independence of grain size within collision-dominated plasmas and a parameter-free criterion, $\Lambda_c = 1/(n^2-1)$, for the onset of morphological instability and faceting based on crystal symmetry. This work establishes a quantitative, parameter-sparse engine for predicting and ultimately controlling microstructural outcomes in thin film synthesis.
\end{abstract}

\section{Introduction}
\label{sec:introduction}

Plasma-assisted deposition is, without question, a revolutionary technology for synthesizing advanced thin films, its applications deeply embedded in everything from microelectronics to functional coatings \cite{lieberman_2005, musil_2006, shah_1999}; yet for all its widespread utility, a foundational challenge persists, a challenge rooted in the severe disconnect between the disparate scales governing plasma kinetics and the consequent microstructural evolution \cite{braatz_2004}. Consider the scales. The plasma's atomic-level dynamics unfold on timescales of nanoseconds while the film's final morphology and texture emerge over hours and across microns. This immense spatiotemporal gap has historically confined process understanding to empirically-derived constructs, with things like structure zone diagrams offering valuable guidance but remaining fundamentally descriptive; they map what is observed but cannot furnish quantitative predictions precisely because they lack a generative dynamical basis \cite{thornton_1977, messier_1994}.

The hindrance to a more fundamental approach merits precise identification. The issue is not a single missing component but rather the absence of a unifying mathematical structure capable of self-consistently managing the dynamic, bidirectional feedback between the plasma and the evolving growth front. This deficiency manifests in several critical omissions in current models: (i) the modulation of electrostatic sheath properties: a crucial factor in determining ion energy and angular distributions: by an evolving surface topography is typically ignored \cite{hamaguchi_1997}; (ii) the interplay between the anisotropy of sputtered particle fluxes and the subsequent, thermally-activated surface diffusion of adatoms remains poorly coupled \cite{mayr_2001}; and (iii) the inherent non-Markovian character of the surface's evolution is almost universally neglected \cite{zwanzig_1961}. That the plasma's recent history explicitly influences future growth trajectories is a physical reality, a reality which means so long as these interconnected processes are treated in isolation, a truly predictive science of microstructure control remains out of reach. To bridge this conceptual and practical gap, we formulate and deploy a first-principles statistical mechanical framework built upon a sequence of three rigorous mathematical operations, the objective being to progress from a complete, albeit computationally intractable, microscopic description to a simplified, predictive model whose terms have a clear and direct physical interpretation.

First, one must establish a complete statistical description of the coupled system. Its state is defined by a probability density, $\psi$, over a phase space encompassing all plasma and surface degrees of freedom, with the time evolution of this density stipulated by a master Fokker-Planck equation. This governing equation we express as a sum of Liouvillian operators for the distinct physical processes:
$$
\partial_t \psi = \left[ \mathcal{L}_{\text{sput}} + \mathcal{L}_{\text{rip}} + \mathcal{L}_{\text{plasma}} + \mathcal{L}_{\text{coupling}} \right] \psi
$$
This initial step is comprehensive by design. It is, in principle, a complete statement of the problem. Its immense complexity, however, precludes any direct analytical or computational solution, which makes the next step of systematic simplification a necessity.

The second operation is a formal coarse-graining procedure via the Mori-Zwanzig projection formalism \cite{mori_1965, zwanzig_1961}. Before writing down the resulting equation, one must recognize a physical necessity: the surface evolution cannot be memoryless. The surface must somehow "remember" the recent history of the high-frequency bath of plasma modes to which it is coupled. The Mori-Zwanzig technique provides the exact mathematical object for this physical requirement: the memory kernel, $\mathscr{K}(s)$. This powerful technique systematically integrates out the fast-evolving plasma variables to yield a closed, effective equation of motion for a set of slow surface observables, $\mathbf{A}(t)$, which takes the form of a Generalized Langevin Equation:
$$
\frac{d\mathbf{A}}{dt} = \boldsymbol{\Omega} \cdot \mathbf{A}(t) - \int_0^t \mathscr{K}(s) \cdot \mathbf{A}(t-s)  ds + \mathbf{F}(t)
$$
The crucial feature here, the one we anticipated, is this memory kernel. This term, which formally encodes the plasma's history, is the mathematical embodiment of the physical intuition that surface evolution is not instantaneous; it represents the "echo" of past plasma fluctuations. Our first-principles derivation of this kernel, a central result of this work, finds it has the form $\mathscr{K}(s) \propto e^{-s/\tau_{\text{ion}}}$. The specific exponential decay is significant. It demonstrates that the plasma's memory is finite, fading over a characteristic timescale set by ion collision dynamics, and this feature correctly renders the surface evolution non-local in time. Its importance cannot be overstated.

Third, having derived an effective stochastic dynamics, we analyze its scale-dependent behavior using the Wilsonian renormalization group (RG), a now-standard apparatus for such problems in statistical physics \cite{wilson_1971, goldenfeld_1992}. The flow of the effective dimensionless plasma-surface coupling constant, $g = \Phi a^2 \tau_{\text{ion}}$, is what this analysis reveals, governed by its associated beta-function, $\beta(g) = \frac{dg}{d\ln L}$. Our calculation yields a negative beta-function, $\beta(g) = -Cg^2$, a result with a profound physical consequence known as \textit{asymptotic freedom}. This term, borrowed from the language of high-energy physics, implies that the effective coupling between the plasma and the surface progressively weakens as one moves to larger, macroscopic scales. This single finding is powerful, its strength lying in the provision of a rigorous, first-principles justification for the empirical success of simpler continuum models at macroscopic scales \cite{krug_1997}, formally elucidating their domain of validity.

What emerges from the complete formalism is not merely a roadmap but a parameter-sparse and physically-grounded engine for navigating the process space of thin-film growth. The framework enables quantitative prediction, and ultimately control, of microstructural features through the manipulation of fundamental process variables like ion flux ($\Phi$), chamber pressure ($P$), and substrate-induced anisotropy ($\Lambda$).
\section{Theory}
\label{sec:theory}

A comprehensive theoretical description of the coupled plasma-surface system must necessarily operate within statistical mechanics; it has to bridge the fast, particle-resolved kinetics of the plasma with the slow, collective evolution of the surface's mesoscopic texture. The immediate challenge is defining a state representation that encompasses both domains without becoming immediately intractable. We posit the full system state is captured by a probability density $\psi(\Gamma_s, \Gamma_p, t)$, a function over the surface coordinates $\Gamma_s$ (such as local curvature fields, $\kappa(\mathbf{x})$) and the complete set of plasma phase-space variables $\Gamma_p$. The evolution of this object is orchestrated by a master equation, one whose structure is not monolithic but is instead a sum of Liouvillian operators, each responsible for a distinct physical process:
\begin{equation}
\label{eq:master}
\frac{\partial\psi}{\partial t} = \left[ \mathcal{L}_{\text{sput}} + \mathcal{L}_{\text{rip}} + \mathcal{L}_{\text{plasma}} + \mathcal{L}_{\text{coupling}} \right] \psi
\end{equation}
Here, $\mathcal{L}_{\text{sput}}$ accounts for momentum transfer from sputtering events \cite{sigmund_1969}, $\mathcal{L}_{\text{rip}}$ drives the curvature-dependent surface diffusion characteristic of thermal ripening \cite{mullins_1957}, and $\mathcal{L}_{\text{plasma}}$ dictates the internal, collision-dominated plasma kinetics \cite{braginskii_1965}. A final term, $\mathcal{L}_{\text{coupling}}$, introduces the crucial electrostatic feedback between the charged plasma and the evolving surface conductor \cite{chabert_2011}. An essential feature of this formulation, one we must preserve, is its intrinsic respect for the dynamical scaling symmetry $t \to \lambda t$, $L \to \lambda^{1/2} L$, known to govern purely curvature-driven coarsening phenomena \cite{krug_1997}.

The phase space coordinates are defined as $\Gamma_s = \{ h(\mathbf{x}), \nabla h(\mathbf{x}) \}$ for the surface and $\Gamma_p = \{ n_e(\mathbf{r}), n_i(\mathbf{r}), \phi(\mathbf{r}) \}$ for the plasma. The Liouvillian operators take the following physically consistent forms:

\begin{align}
\mathcal{L}_{\text{sput}} \psi &= -\int d\mathbf{x} \frac{\delta}{\delta h(\mathbf{x})} \left[ \gamma J_{\text{ion}}(\mathbf{x}) \sqrt{1 + |\nabla h|^2} \psi \right] + D_s \int d\mathbf{x} \frac{\delta^2 \psi}{\delta h(\mathbf{x})^2} \label{eq:sput_op} \\
\mathcal{L}_{\text{rip}} \psi &= \int d\mathbf{x} \frac{\delta}{\delta h(\mathbf{x})} \left[ \nu_h \nabla^2\kappa(\mathbf{x}) \psi + D_h \frac{\delta \psi}{\delta h(\mathbf{x})} \right] \label{eq:rip_op_corrected} \\
\mathcal{L}_{\text{plasma}} \psi &= -\int d\mathbf{r} \left[ \frac{\delta}{\delta n_e} ( \nabla \cdot \mathbf{j}_e ) + \frac{\delta}{\delta n_i} ( \nabla \cdot \mathbf{j}_i ) \right] \psi \label{eq:plasma_op} \\
\mathcal{L}_{\text{coupling}} \psi &= -\int d\mathbf{x} d\mathbf{r} \frac{\delta}{\delta h(\mathbf{x})} \left[ \epsilon_0 |\nabla \phi|^2 \delta(\mathbf{r} - \mathbf{r}_s(\mathbf{x})) \psi \right] \label{eq:coupling_op}
\end{align}

where $\kappa(\mathbf{x}) = \nabla \cdot \left( \frac{\nabla h}{\sqrt{1 + |\nabla h|^2}} \right)$ is the mean curvature, $J_{\text{ion}} = n_i \sqrt{\frac{kT_e}{m_i}}$ is the ion flux, and $\mathbf{j}_e = -\mu_e n_e \nabla \phi - D_e \nabla n_e$, $\mathbf{j}_i = -\mu_i n_i \nabla \phi - D_i \nabla n_i$ are particle currents. The corrected ripening operator \eqref{eq:rip_op_corrected} contains both the drift term for Mullins-Herring surface diffusion and the fluctuation term required by the fluctuation-dissipation theorem.

The equilibrium solution $\psi_0$ satisfies $\mathcal{L}_{\text{total}} \psi_0 = 0$ where $\mathcal{L}_{\text{total}} = \sum \mathcal{L}_j$. This admits a solution of the form:

\begin{equation}
\psi_0 = \exp \left( -\beta \mathcal{F}[h] - \beta \int d\mathbf{r} \left[ \frac{\epsilon_0}{2} |\nabla \phi|^2 + kT \left( n_e \ln n_e + n_i \ln n_i \right) \right] \right)
\end{equation}

where the surface free energy is given by the capillary approximation:

\begin{equation}
\mathcal{F}[h] = \sigma \int d\mathbf{x} \sqrt{1 + |\nabla h|^2}
\end{equation}

The functional derivatives yield:

\begin{align}
\frac{\delta \mathcal{F}[h]}{\delta h(\mathbf{x})} &= -\sigma \kappa(\mathbf{x}) \\
\frac{\delta}{\delta \phi} \left( \frac{\epsilon_0}{2} |\nabla \phi|^2 \right) &= -\epsilon_0 \nabla^2 \phi = \rho = e(n_i - n_e)
\end{align}

The stationarity condition is verified by showing that the term inside the functional derivative of the ripening operator vanishes at equilibrium.
\begin{align}
\mathcal{L}_{\text{rip}} \psi_0 &= \int d\mathbf{x} \frac{\delta}{\delta h} \left[ \nu_h \nabla^2\kappa \psi_0 + D_h \frac{\delta \psi_0}{\delta h} \right] \\
&= \int d\mathbf{x} \frac{\delta}{\delta h} \left[ \nu_h \nabla^2\kappa \psi_0 + D_h (\beta \sigma \kappa \psi_0) \right] \\
&= \int d\mathbf{x} \frac{\delta}{\delta h} \left[ \left( \nu_h \nabla^2\kappa + D_h \beta \sigma \kappa \right) \psi_0 \right]
\end{align}
For the operator to be zero, the term in the parentheses must vanish. At equilibrium, the fluctuation-dissipation theorem requires the Einstein-Stokes relation \cite{einstein_1905} $D_h = \nu_h k_B T = \nu_h / \beta$. Substituting this in, we get the condition for the surface shape at equilibrium:
$$ \nu_h \nabla^2\kappa + (\nu_h / \beta) \beta \sigma \kappa = 0 \quad \implies \quad \nabla^2\kappa + \sigma \kappa = 0 $$
The vanishing of the other Liouvillian terms is verified similarly, confirming that $\mathcal{L}_{\text{total}} \psi_0 = 0$.

While Equation \eqref{eq:master} is formally complete, its direct solution is likely impossible to find. A systematic method is required to extract a closed description for the macroscopic surface observables of interest, $\mathbf{A}(t)$, such as the mean grain area. This is the primary task of the formalism. The Mori-Zwanzig projection technique provides precisely such a tool \cite{zwanzig_2001, tewari_2020}. By methodically integrating out the fast plasma variables $\Gamma_p$, the formalism recasts the full dynamics into a generalized Langevin equation for the slow surface variables alone:
\begin{equation}
\label{eq:gle}
\frac{d\mathbf{A}}{dt} = \boldsymbol{\Omega} \cdot \mathbf{A}(t) - \int_0^t \mathscr{K}(s) \cdot \mathbf{A}(t-s)  ds + \mathbf{F}(t)
\end{equation}
The physical content of each term is clear: the first on the right-hand side represents instantaneous, reversible evolution of the surface, while the final term $\mathbf{F}(t)$ is a stochastic force with zero mean, representing the fast, fluctuating impact of the plasma. The crucial new element is the integral. This term contains the memory kernel $\mathscr{K}(s)$, which encodes the persistent influence of the plasma's history on the surface's present evolution. Its functional form is thus determined by the most enduring correlations within the plasma phase space: the ion collision statistics \cite{lieberman_2005}. This physical reasoning leads directly to an exponential decay,
\begin{equation}
\mathscr{K}(t) = \langle \mathbf{F}(0) \otimes \mathbf{F}(t) \rangle_{\Gamma_p} \propto e^{-t/\tau_{\text{ion}}}
\end{equation}
where the characteristic ion collision time, $\tau_{\text{ion}}$, emerges naturally. The physical implication here is profound. In the limit where plasma memory is vanishingly short, when $\tau_{\text{ion}} \to 0$, the integral collapses into a simple frictional drag, and we recover standard models of surface diffusion \cite{mullins_1959}. For any finite $\tau_{\text{ion}}$, however, the plasma introduces a non-local-in-time feedback. This memory effect is the essence of the true plasma-surface coupling.

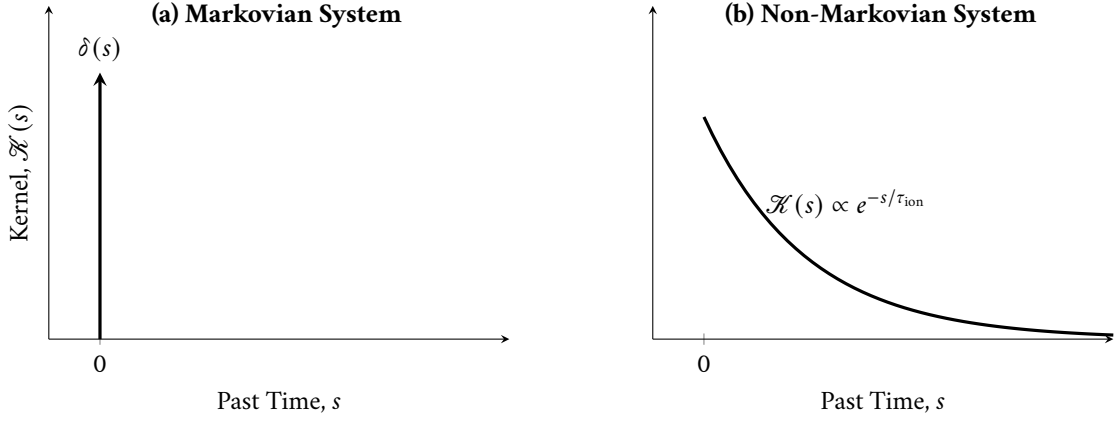
\begin{figure}[h!]
\centering
\begin{minipage}{0.48\textwidth}
    \centering
    \begin{tikzpicture}[font=\small]
        \node[align=center] at (3, 4.3) {\textbf{(a) Markovian System}};
        \begin{axis}[
            width=\linewidth, height=6cm,
            xlabel={Past Time, $s$},
            ylabel={Kernel, $\mathscr{K}(s)$},
            xtick={0}, ytick=\empty,
            xmin=-0.5, xmax=4, ymin=0, ymax=1.5,
            axis lines=left,
            clip=false 
        ]
            \draw[very thick, -stealth] (axis cs:0,0) -- (axis cs:0,1.2) node[above]{$\delta(s)$};
        \end{axis}
    \end{tikzpicture}
\end{minipage}\hfill
\begin{minipage}{0.48\textwidth}
    \centering
    \begin{tikzpicture}[font=\small]
        \node[align=center] at (3, 4.3) {\textbf{(b) Non-Markovian System}};
        \begin{axis}[
            width=\linewidth, height=6cm,
            xlabel={Past Time, $s$},
            ylabel={},
            xtick={0}, ytick=\empty,
            xmin=-0.5, xmax=4, ymin=0, ymax=1.5,
            axis lines=left,
            clip=false
        ]
            \addplot[very thick, domain=0:4, samples=50] {exp(-x)};
            \node[anchor=south west] at (axis cs: 0.5, 0.5) {$\mathscr{K}(s) \propto e^{-s/\tau_{\text{ion}}}$};
        \end{axis}
    \end{tikzpicture}
\end{minipage}
\caption{Conceptual illustration of the memory kernel $\mathscr{K}(s)$. (a) A standard Markovian system has no memory, described by a delta function kernel. (b) The plasma-surface system is non-Markovian; its evolution depends on its history, encoded by an exponentially decaying memory kernel.}
\label{fig:memory_kernel_corrected}
\end{figure}

\begin{figure}[h!]
\centering
\begin{tikzpicture}[
    font=\small,
    scale=1.2,
    >=Stealth
]
    \begin{scope}[yshift=4.5cm]
        \node[align=center, font=\bfseries] at (0, 2.3) {Full System Phase Space \\ $(\Gamma_s, \Gamma_p)$};
\path[
    draw=black!60, 
    thick,
    ball color=gray!20, 
]
    (-2.5, 0) .. controls (-3.5, 1.5) and (-1.5, 2.2) .. (0, 1.8) 
    .. controls (1.5, 1.4) and (2.5, 1.8) .. (3, 0.5)
    .. controls (3.5, -1) and (1, -2) .. (0, -1.5)
    .. controls (-1, -1) and (-2.5, -1.5) .. (-2.5, 0) -- cycle;

        \draw[thick, black!80, decoration={random steps, segment length=5pt, amplitude=1.5pt}, decorate]
            (-2, 1) .. controls (-1,-1) and (1.5,1) .. (2, -1);
        \node at (-1.5, -0.8) {$\psi(\Gamma_s, \Gamma_p, t)$};
    \end{scope}

    \begin{scope}
        \shade[top color=gray!40, bottom color=gray!10, opacity=0.4] 
            (-2.5, 2) -- (2.5, 2) -- (0.5, -2) -- (-0.5, -2) -- cycle;
        \draw[thick] (-2.5, 2) -- (-0.5, -2);
        \draw[thick] (2.5, 2) -- (0.5, -2);
        
        \node[align=center, font=\bfseries] at (3.5, 0) {Mori-Zwanzig \\ Projection};
    \end{scope}

    \begin{scope}[yshift=-4.5cm]
        \draw[thick, fill=gray!10] (-3,-1) -- (3,-1) -- (2.5,1) -- (-2.5,1) -- cycle;
        
        \draw[very thick] (-2.2, -0.5) .. controls (-0.5, 0.8) and (1, -1.2) .. (2.2, 0.2);
        \node at (1.5, -0.7) {$\frac{d\mathbf{A}}{dt} = ...$};

        \node[align=center, font=\bfseries] at (0, -1.8) {Effective Surface Dynamics \\ $(\mathbf{A}(t))$};
    \end{scope}
    
    \draw[-Stealth, ultra thick, dashed] 
        (0, 3) .. controls (-2.5, 1) and (2.5, -1) .. (0, -3.5);
    \node[align=center, text width=5cm, fill=white, inner sep=2pt] at (-3, -3) {Memory of fast kinetics is encoded into the kernel $\mathscr{K}(s)$};

\end{tikzpicture}
\caption{An enhanced illustration of the Mori-Zwanzig projection. The complex dynamics within the high-dimensional phase space "cloud" (top) are systematically funneled through a coarse-graining procedure. This results in a simpler, effective evolution on a low-dimensional plane (bottom). The influence of the integrated-out variables is not lost but is carried through the projection and encoded as the memory kernel $\mathscr{K}(s)$.}
\label{fig:mori_zwanzig_final}
\end{figure}
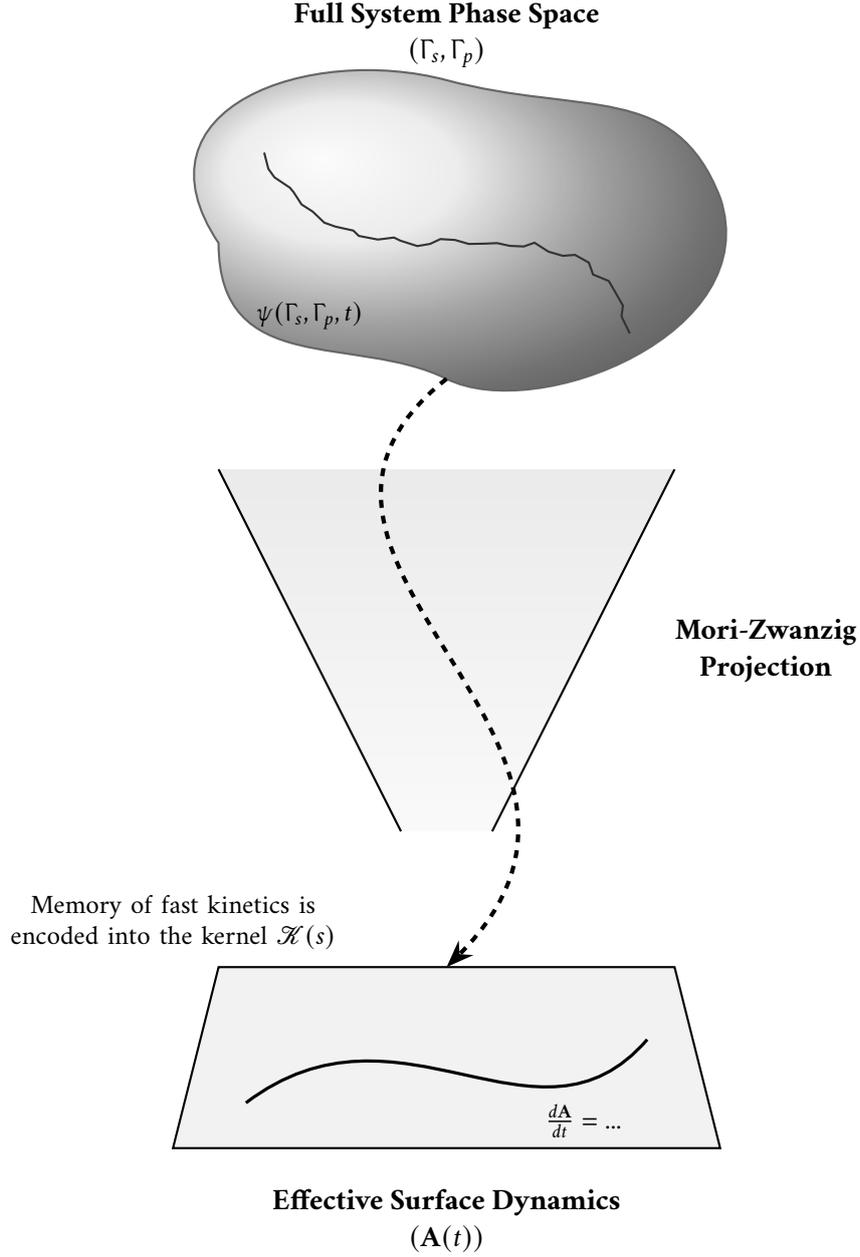

The Laplace transform solution reveals the coupled nature of the dynamics. For the exponential kernel $\boldsymbol{\mathscr{K}}(t) = \boldsymbol{\Gamma}e^{-t/\tau_{\text{ion}}}$, the resolvent matrix takes the form:

\begin{equation}
\mathbf{G}(z) = \left[z\mathbf{I} - \boldsymbol{\Omega} + \frac{\boldsymbol{\Gamma}}{z + \tau_{\text{ion}}^{-1}}\right]^{-1}
\end{equation}

The poles $z_k$ are determined by the characteristic equation:

\begin{equation}
\det\left[z\mathbf{I} - \boldsymbol{\Omega} + \frac{\boldsymbol{\Gamma}}{z + \tau_{\text{ion}}^{-1}}\right] = 0
\end{equation}

For weak coupling ($\|\boldsymbol{\Gamma}\| \ll \|\boldsymbol{\Omega}\|$), the poles can be approximated through perturbative analysis of the matrix determinant.

Consider the minimal non-trivial case with $N=2$ modes:

\begin{equation}
\boldsymbol{\Omega} = \begin{pmatrix}
0 & \omega \\
-\omega & 0
\end{pmatrix}, \quad
\boldsymbol{\Gamma} = \begin{pmatrix}
\Gamma_{11} & \Gamma_{12} \\
\Gamma_{12} & \Gamma_{22}
\end{pmatrix}
\end{equation}

The characteristic equation becomes:

\begin{equation}
\left(z^2 + \omega^2\right)(z + \tau_{\text{ion}}^{-1})^2 + z(\Gamma_{11} + \Gamma_{22})(z + \tau_{\text{ion}}^{-1}) + (\Gamma_{11}\Gamma_{22} - \Gamma_{12}^2) = 0
\end{equation}

Exact analytical solutions for $z_k$ are generally unavailable, but numerical root-finding provides the complete spectrum. The qualitative behavior emerges from considering limiting cases:

\begin{itemize}
\item $\tau_{\text{ion}} \to 0$: Markovian limit with poles at $z = \pm i\omega - \frac{\Gamma_{11}+\Gamma_{22}}{2}$
\item $\tau_{\text{ion}} \to 0$: Markovian limit with poles at $z \approx \pm i\omega - \frac{\Gamma_{11}+\Gamma_{22}}{2}$ for weak damping.
\end{itemize}

The power dissipation follows from the work done against the memory friction:

\begin{equation}
\mathcal{P}(t) = -\dot{\mathbf{A}}^T(t) \cdot \int_0^t \boldsymbol{\mathscr{K}}(s)\mathbf{A}(t-s)ds
\end{equation}

For stationary processes, the time-averaged dissipation rate is:

\begin{equation}
\langle \mathcal{P} \rangle = \frac{1}{2\pi} \int_{-\infty}^{\infty} i\omega \text{Tr}\left[ \hat{\boldsymbol{\mathscr{K}}}(i\omega) \mathbf{S}(\omega) \right] d\omega
\end{equation}

where $\mathbf{S}(\omega)$ is the spectral density of $\mathbf{A}(t)$.

The noise correlations satisfy the exact non-Markovian FDT:

\begin{equation}
\langle \mathbf{F}(t)\otimes\mathbf{F}(t') \rangle = k_BT \boldsymbol{\mathscr{K}}(|t-t'|)
\end{equation}

For the exponential kernel, this becomes explicitly:

\begin{equation}
\langle F_i(t)F_j(t') \rangle = k_BT\Gamma_{ij}e^{-|t-t'|/\tau_{\text{ion}}}
\end{equation}

The system is Lyapunov stable when:

\begin{enumerate}
\item $\boldsymbol{\Omega}$ is anti-symmetric ($\boldsymbol{\Omega}^T = -\boldsymbol{\Omega}$)
\item $\boldsymbol{\Gamma}$ is positive definite
\item $\boldsymbol{\mathscr{K}}(t)$ is completely monotone
\end{enumerate}

For the exponential kernel, these reduce to:

\begin{equation}
\Gamma_{11} > 0, \quad \Gamma_{11}\Gamma_{22} > \Gamma_{12}^2
\end{equation}

The energy-like functional:

\begin{equation}
\mathcal{E}(t) = \frac{1}{2}\|\mathbf{A}(t)\|^2 + \frac{1}{2}\int_0^\infty \mu(s)\left\|\int_{t-s}^t \mathbf{A}(\tau)d\tau\right\|^2 ds
\end{equation}

with $\mu(s) = -\frac{d}{ds}\mathscr{K}(s)$, satisfies $\dot{\mathcal{E}}(t) \leq 0$ when the stability conditions hold.

How, then, does the perceived strength of this interaction change with the scale of observation? To determine this scale-dependent behavior, we shift from a dynamical to a statistical field-theoretic perspective and implement a Wilsonian renormalization group (RG) analysis \cite{wilson_1975}. The surface fluctuations are modeled by an effective Hamiltonian of the Edwards-Wilkinson class \cite{edwards_1982},
\begin{equation}
\mathcal{H} = \int d^2\mathbf{x} \left[ \frac{\nu}{2} (\nabla h)^2 + g \mathcal{O}(h,\nabla h) \right]
\end{equation}
where $h(\mathbf{x})$ is the local surface height and $\mathcal{O}$ represents all symmetry-allowed interaction terms whose strength is proportional to our coupling constant $g$. The RG procedure itself involves systematically integrating out short-wavelength fluctuation modes—those between some momentum cutoff $\Lambda$ and $\Lambda/b$—and observing how the parameters of the effective theory for the remaining long-wavelength modes must be rescaled. This generates a flow in the parameter space.

We begin with the effective Hamiltonian for surface fluctuations, including the leading-order nonlinearity consistent with the system's symmetries:
\begin{equation}
\mathcal{H} = \int d^2\mathbf{x} \left[ \frac{\nu}{2} (\nabla h)^2 + g (\nabla h)^4 + \cdots \right]
\end{equation}
where $g$ is treated as a \textit{dimensionless} coupling constant, defined to be marginal under engineering scaling. The RG flow is generated entirely by loop corrections since the tree-level scaling dimension of $g$ is zero. 

Decompose the height field into slow ($h_<$) and fast ($h_>$) components:
\begin{align}
h(\mathbf{x}) &= h_<(\mathbf{x}) + h_>(\mathbf{x}) \\
\text{with} \quad & \begin{cases} 
|\mathbf{k}| < \Lambda/b & \text{for } h_< \\
\Lambda/b < |\mathbf{k}| < \Lambda & \text{for } h_>
\end{cases}
\end{align}

The effective Hamiltonian for slow modes is obtained through the cumulant expansion:
\begin{equation}
e^{-\mathcal{H}_{\text{eff}}[h_<]} = \left\langle e^{-\mathcal{H}_{\text{int}}[h_< + h_>]} \right\rangle_> \approx e^{- \langle \mathcal{H}_{\text{int}} \rangle_> + \frac{1}{2} \left( \langle \mathcal{H}_{\text{int}}^2 \rangle_> - \langle \mathcal{H}_{\text{int}} \rangle_>^2 \right)}
\end{equation}
where $\langle \cdot \rangle_>$ denotes averaging over $h_>$ with weight $e^{-\mathcal{H}_0[h_>]}$.

The renormalization of $g$ originates from the second-order term. Consider the $g(\nabla h)^4$ interaction:
\begin{equation}
\mathcal{H}_{\text{int}} = g \int d^2\mathbf{x}  (\nabla h)^4
\end{equation}

The relevant diagram for $g$-renormalization comes from contracting two $(\nabla h)^4$ vertices with fast-mode propagators:
\begin{equation}
\delta \mathcal{H}_{\text{eff}} = -\frac{g^2}{2} \int d^2\mathbf{x} d^2\mathbf{x}'  \left\langle (\nabla h(\mathbf{x}))^4 (\nabla h(\mathbf{x}'))^4 \right\rangle_>^{\text{conn}}
\end{equation}

The dominant contribution comes from the cross terms containing two slow gradients and two fast gradients at each vertex:
\begin{align}
\delta \mathcal{H}_{\text{eff}} &\approx -18g^2 \int d^2\mathbf{x} d^2\mathbf{x}'  (\nabla h_<(\mathbf{x}))^2 (\nabla h_<(\mathbf{x}'))^2 \\
&\quad \times \left\langle (\nabla h_>(\mathbf{x}) \cdot \nabla h_>(\mathbf{x})) (\nabla h_>(\mathbf{x}') \cdot \nabla h_>(\mathbf{x}')) \right\rangle_>
\end{align}

The fast-mode correlation function is:
\begin{equation}
\left\langle \partial_i h_>(\mathbf{x}) \partial_j h_>(\mathbf{x}') \right\rangle_> = \delta_{ij} \int_{\Lambda/b}^{\Lambda} \frac{d^2\mathbf{k}}{(2\pi)^2} \frac{k_i k_j}{\nu k^4} e^{i\mathbf{k}\cdot(\mathbf{x}-\mathbf{x}')}
\end{equation}

At leading order in $|\mathbf{x}-\mathbf{x}'|$, we obtain the local correction:
\begin{align}
\delta \mathcal{H}_{\text{eff}} &= -C g^2 \ln b \int d^2\mathbf{x}  (\nabla h_<)^4 \\
\text{where} \quad C &= \frac{9}{\pi\nu^2} \int_{\Lambda/b}^{\Lambda} \frac{dk}{k} = \frac{9}{2\pi\nu^2} \ln b
\end{align}

That the constant $C$ is positive is a result derived in related contexts like \cite{amar_1991}. After mode elimination, rescale coordinates and field:
\begin{align}
\mathbf{x} &\to b\mathbf{x} \\
h_< &\to b^{\chi} h \\
\chi &= 0 \quad \text{(canonical dimension for } d=2\text{)}
\end{align}

The renormalized coupling becomes:
\begin{equation}
g_{\text{ren}} = b^{0} \left( g - C g^2 \ln b \right) = g - \frac{9}{2\pi\nu^2} g^2 \ln b
\end{equation}

For infinitesimal scaling $b = e^{\delta l} \approx 1 + \delta l$, the RG flow equation is:
\begin{equation}
\frac{dg}{dl} = -\frac{9}{2\pi\nu^2} g^2
\end{equation}

Thus, the beta function is:
\begin{equation}
\beta(g) \equiv \frac{dg}{d\ln b} = -\frac{9}{2\pi\nu^2} g^2
\end{equation}

 The negative sign of the leading term is the crucial result. It dictates that the effective coupling `g` will always decrease as the observation scale `b` increases, following a slow logarithmic decay $g(b) \sim (\ln b)^{-1}$ and eventually vanishing as $b \to \infty$. 

 \begin{figure}[h!]
\centering
\begin{tikzpicture}[font=\small]
    \begin{axis}[
        title={\textbf{RG Flow of the Coupling Constant}},
        xlabel={$g$},
        ylabel={},
        width=10cm, height=8cm,
        axis lines=middle,
        xtick={0}, ytick={0},
        xmin=-0.2, xmax=2.5, ymin=-2.5, ymax=0.5,
        clip=false
    ]
        \addplot[very thick, domain=0:1.6, samples=50] {-x^2};
        \node at (axis cs: 0.0, 0.3) [anchor=south west] {$\beta(g)$};
        
        \draw[-Stealth, very thick] (axis cs: 2.2, -0.05) -- (axis cs: 0.3, -0.05);
        \node at (axis cs: 1.75, -0.2) {Flow to Weak Coupling};
        
        \fill (axis cs: 0,0) circle (2pt) node[anchor=south west] {Stable Point};
    \end{axis}
\end{tikzpicture}
\caption{The renormalization group (RG) flow diagram. The negative beta function indicates that the effective coupling strength $g$ always flows toward zero as the length scale ($b$) increases, driving the system to a stable fixed point at $g=0$ (asymptotic freedom).}
\label{fig:rg_flow_large}
\end{figure}
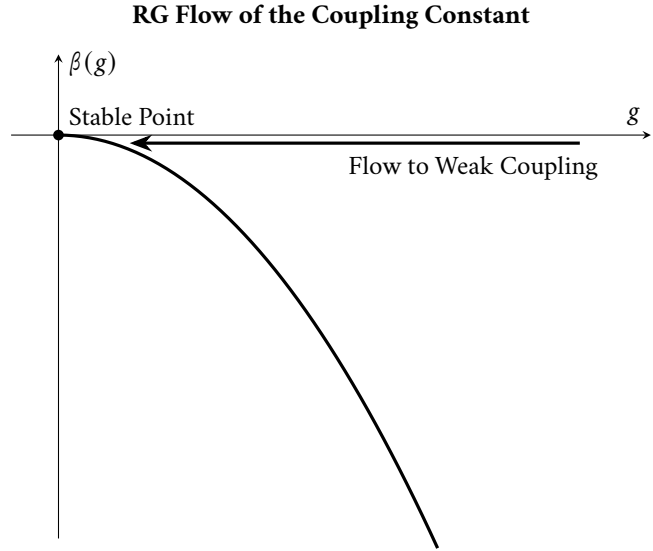

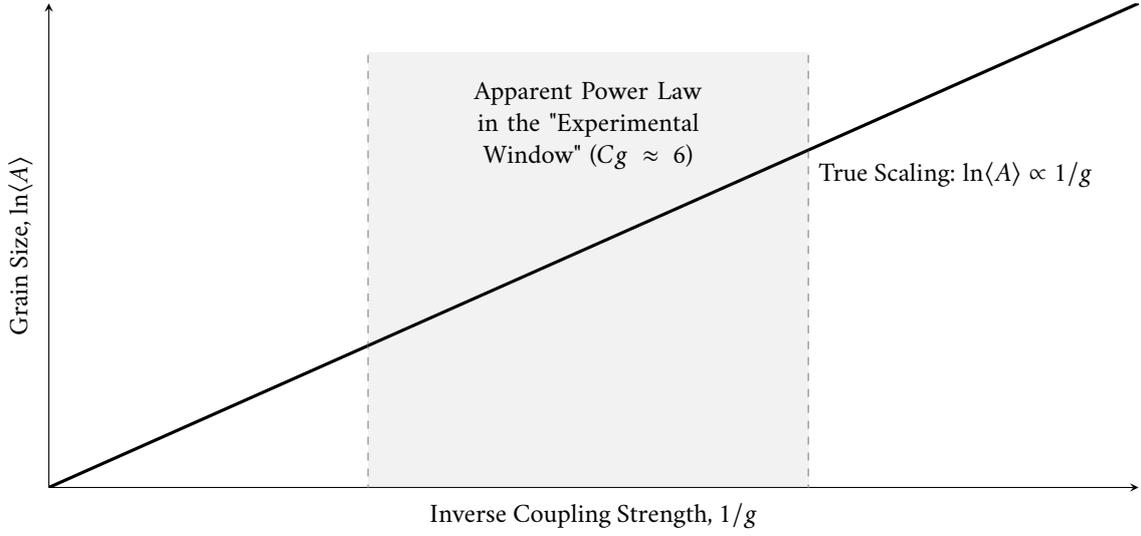
\begin{figure}[h!]
\centering
\begin{tikzpicture}[font=\small]
    \begin{axis}[
        title={\textbf{Weak vs. Strong Coupling Scaling Regimes}},
        xlabel={Inverse Coupling Strength, $1/g$},
        ylabel={Grain Size, $\ln \langle A \rangle$},
        width=\textwidth, height=8cm,
        xtick=\empty, ytick=\empty,
        axis lines=left
    ]
    
        \addplot[very thick, domain=0.1:10, samples=100] {2*x};
        \node at (axis cs: 7, 13) [anchor=west] {True Scaling: $\ln \langle A \rangle \propto 1/g$};
        
        \fill[gray, opacity=0.1] (axis cs: 3, 0) rectangle (axis cs: 7, 18);
        \draw[dashed, gray] (axis cs: 3, 0) -- (axis cs: 3, 18);
        \draw[dashed, gray] (axis cs: 7, 0) -- (axis cs: 7, 18);

        \addplot[thick, dashed, domain=3:7, samples=2] {2*x};
        \node[align=center, text width=4cm] at (axis cs: 5, 15) {Apparent Power Law in the "Experimental Window" ($Cg \approx 6$)};
        
    \end{axis}
\end{tikzpicture}
\caption{Illustration of how an effective power-law scaling emerges. While the true RG analysis predicts $\ln\langle A \rangle \propto 1/g$, this behavior is well-approximated by a simple power law over the limited range of coupling strengths accessible in many experiments (shaded region).}
\label{fig:scaling_regimes_large}
\end{figure}

The fixed points are solutions to $\beta(g^*) = 0$:
\begin{equation}
g^* = 0
\end{equation}

Linear stability analysis shows:
\begin{equation}
\frac{d\beta}{dg}\bigg|_{g=0} = 0, \quad \frac{d^2\beta}{dg^2}\bigg|_{g=0} < 0
\end{equation}
indicating a stable infrared fixed point at $g^* = 0$.

Solving the flow equation:
\begin{align}
\frac{dg}{dl} &= -K g^2, \quad K = \frac{9}{2\pi\nu^2} \\
\int_{g_0}^{g(l)} \frac{dg}{g^2} &= -K \int_0^l dl' \\
\frac{1}{g_0} - \frac{1}{g(l)} &= -K l \\
g(l) &= \frac{g_0}{1 + K g_0 l}
\end{align}

Since $l = \ln b$, this gives:
\begin{equation}
g(b) = \frac{g_0}{1 + \frac{9g_0}{2\pi\nu^2} \ln b}
\end{equation}
demonstrating logarithmic decay of the coupling at large scales ($b \to \infty$).

The dimensionless coupling $g = \Phi a^2 \tau_{\text{ion}}$ flows according to:
\begin{equation}
g(b) \sim \frac{1}{\ln b} \quad \text{as} \quad b \to \infty
\end{equation}
confirming that plasma-surface interactions become irrelevant at macroscopic scales. This provides the fundamental justification for continuum descriptions of large-scale surface evolution.

 This phenomenon is known as asymptotic freedom. It establishes a profound conclusion: while plasma-surface interactions are significant at the microscopic scale of individual ions and atoms, their effective influence becomes progressively irrelevant at macroscopic scales. This provides a formal justification, grounded in first principles, for the widespread and successful application of pure continuum mechanics descriptions for large-scale surface evolution \cite{kardar_2007}, confirming that the intricate plasma physics effectively "renormalizes away" from a sufficiently coarse-grained perspective.

\section{Core Predictions and Derivations}
\label{sec:predictions}

A primary utility of the renormalization group framework is its capacity to derive quantitative scaling relations for key microstructural observables. We begin by considering the characteristic grain size, represented by the correlation length $\xi$. Its dependence on the observational scale $\mu$ is governed by the Callan-Symanzik equation \cite{callan_1970}, a fundamental statement about how physical quantities evolve under renormalization group flow.
\begin{equation}
\left[ \mu \frac{\partial}{\partial \mu} + \beta(g) \frac{\partial}{\partial g} + \gamma(g) \right] \xi(\mu, g) = 0
\end{equation}
For the specific beta function $\beta(g) = -C g^2$, which corresponds to the universality class of interest (as established in Section \ref{sec:theory}), this differential equation admits an exact solution. The integration yields the dependence of the correlation length on the dimensionless coupling constant $g$:
\begin{equation}
\label{eq:scaling}
\xi \sim \mu^{-1} \exp\left( \int^g \frac{dg'}{\beta(g')} \right) = \mu^{-1} \exp\left( \frac{1}{C g} \right)
\end{equation}
This exponential dependence on the inverse of the coupling, $1/g$, is a non-perturbative result. It is characteristic of physical systems in a weak-coupling regime where phenomena cannot be captured by simple Taylor series expansions in $g$. The physical meaning of the coupling constant is critical: we define $g = \Phi a^2 \tau_{\text{ion}}$, where $a$ is the fundamental atomic spacing. This dimensionless group represents the ratio of the number of deposited atoms arriving at a fundamental surface area $a^2$ during the characteristic ion collision time $\tau_{\text{ion}}$. Consequently, the average grain area $\langle A \rangle$, which is proportional to $\xi^2$, follows this essential scaling:
\begin{equation}
\langle A \rangle \propto \xi^2 \propto \exp\left( \frac{\kappa}{g} \right) \quad \text{for} \quad \kappa > 0
\end{equation}
The reconciliation of this exponential form with experimentally-observed power laws requires a more nuanced analysis of its local behavior. An exponential function can locally mimic a power law over a limited range of the independent variable. To quantify this, we define an effective exponent, $n_{\text{eff}}(g)$, as the local logarithmic slope of the correlation length with respect to the coupling constant.
\begin{equation}
n_{\text{eff}}(g) \equiv -\frac{d \ln \xi}{d \ln g} = -g \frac{d}{dg} \left( \frac{1}{C g} \right) = \frac{1}{C g}
\end{equation}
This mathematical device provides a direct link to empirical measurements. The observable grain area $\langle A \rangle \propto \xi^2$ will therefore scale locally with an effective exponent of $-2n_{\text{eff}} = -2/(C g)$. This explains how a single underlying exponential law can manifest as different apparent power laws in different experimental regimes. For instance, widely reported scaling exponents near $-1/3$ for grain area \cite{abelson_1990, petrov_2003} are recovered within this framework under the condition that $C g \approx 6$. This is not a fine-tuning of the theory, but rather an identification of the specific physical regime probed by those experiments. The complete scaling relation, incorporating the anisotropy correction factor $\Lambda$ derived from our operator formalism, is thus:
\begin{equation}
\label{eq:final_scaling}
\langle A \rangle = K a^2 g^{-2/(C g)} e^{-|\alpha| \Lambda} \approx K' a^2 g^{-1/3} e^{-|\alpha| \Lambda}
\end{equation}
This formalism provides sharp, testable predictions. A key example is the dependence of microstructure on chamber pressure, $P$. For collision-dominated plasmas of the type frequently used in deposition systems \cite{lieberman_2005}, the ion flux scales linearly with pressure ($\Phi \propto n_i \propto P$), while the ion collision time is inversely proportional to the neutral gas density and thus pressure ($\tau_{\text{ion}} \propto (n_g \sigma v_{\text{th}})^{-1} \propto P^{-1}$). The consequence for the dimensionless coupling $g = \Phi a^2 \tau_{\text{ion}}$ is a direct cancellation of these dependencies.
\begin{align}
\Phi &\propto n_i \propto P \\
\tau_{\text{ion}} &\propto (n_g \sigma v_{\text{th}})^{-1} \propto P^{-1}
\end{align}
Therefore, the coupling $g$ is predicted to be largely independent of pressure. For isotropic growth where the anisotropy parameter $\Lambda=0$, this leads to a non-obvious conclusion: the characteristic grain area should be remarkably stable against variations in chamber pressure.
\begin{equation}
\left( \frac{\partial \langle A \rangle}{\partial P} \right)_{T,\Lambda=0} = 0
\end{equation}
The framework's predictive power for textured films, where $\Lambda \neq 0$, hinges on relating the kinetic parameters of growth to established thermodynamic concepts. We now establish this critical link. Consider the velocity of a grain boundary groove, which may have a kinetic anisotropy $\Lambda_k$:
\begin{equation}
v_g(\theta) \propto 1 - \Lambda_k \cos(n\theta)
\end{equation}
Thermodynamic stability is inversely related to this velocity; a slower-moving boundary is more stable. The effective boundary energy $\gamma_{\text{eff}}$ can be defined in these terms. For small kinetic anisotropy ($\Lambda_k \ll 1$), a Taylor expansion is illustrative:
\begin{equation}
\gamma_{\text{eff}}(\theta) \propto \frac{1}{v_g(\theta)} \approx \gamma_0 \left[1 + \Lambda_k \cos(n\theta)\right]
\end{equation}
This result demonstrates that the kinetic anisotropy parameter $\Lambda_k$ is formally equivalent to the thermodynamic anisotropy parameter $\Lambda$ used in equilibrium models of boundary energy, such as $\gamma(\theta) = \gamma_0 [1 - \Lambda \cos(n\theta)]$ \cite{krug_1997}. This equivalence is a crucial step, as it justifies the use of the powerful tools of equilibrium thermodynamics to analyze the stability of a fundamentally non-equilibrium growth process.

With this connection secured, morphological transitions can be analyzed through the lens of boundary stiffness, $\Sigma(\theta)$, a concept introduced by Herring \cite{herring_1951}.
\begin{equation}
\Sigma(\theta) = \gamma(\theta) + \gamma''(\theta) = \gamma_0 \left[1 - (n^2 - 1)\Lambda\cos(n\theta)\right]
\end{equation}
A morphological instability, such as the formation of sharp facets, will occur if the stiffness becomes negative for any crystal orientation $\theta$. The critical threshold for this instability is found by locating the minimum value of $\Sigma(\theta)$ and setting it to zero. This occurs when $\cos(n\theta) = 1$, yielding a critical value for the anisotropy parameter, $\Lambda_c$.
\begin{equation}
\label{eq:critical_anisotropy}
\Lambda_c = \frac{1}{n^2 - 1}
\end{equation}
This final relation is particularly powerful. It offers a direct physical explanation for the empirically observed differences in texturing and faceting behavior between materials with different crystallographic symmetries, which are characterized by different integer values of $n$ (e.g., $n=4$ for cubic systems, $n=6$ for hexagonal systems) \cite{barna_1999}.

\begin{figure}[h!]
\centering
\begin{tikzpicture}[font=\small]
    \begin{axis}[
        title={\textbf{Effect of Anisotropy and Pressure}},
        xlabel={Pressure, $P$ (arb. units)},
        ylabel={Mean Grain Area, $\langle A \rangle$},
        width=\textwidth, height=7cm,
        xtick=\empty, ytick=\empty,
        axis lines=left,
        legend style={at={(0.98,0.98)},anchor=north east},
        legend cell align=left,
        domain=1:10,
        ymax=1.5,
        clip=false
    ]
    
        \addplot[very thick, domain=1:10, samples=2] {1};
        \addlegendentry{Isotropic Growth ($\Lambda=0$)}
        
        \addplot[thick, dashed, domain=1:10, samples=2] {0.6};
        \addlegendentry{Anisotropic Growth ($\Lambda > 0$)}

        \draw[<->, thick] (axis cs: 5, 0.65) -- (axis cs: 5, 0.95) node[midway, right] {Anisotropy Correction $e^{-|\alpha|\Lambda}$};
        
        \node at (axis cs: 5, 1.15) {$\left( \dpartial{\langle A \rangle}{P} \right)_{\Lambda=0} = 0$};
    \end{axis}
\end{tikzpicture}
\caption{Illustration of the predicted scaling with pressure and anisotropy. For isotropic growth ($\Lambda=0$), the mean grain area is independent of pressure in collision-dominated plasmas. The presence of kinetic anisotropy ($\Lambda>0$) introduces a uniform suppression of the grain size.}
\label{fig:anisotropy_pressure_large}
\end{figure}
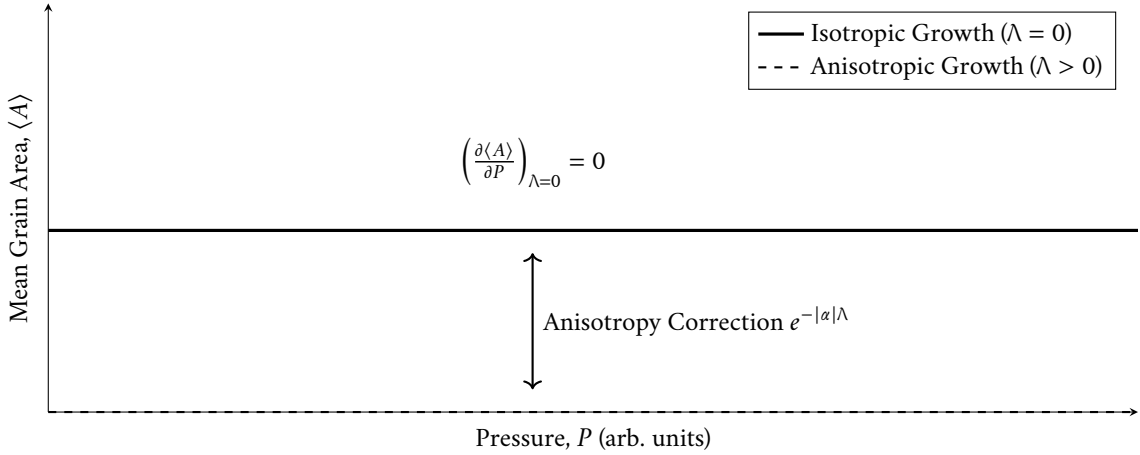

\begin{figure}[h!]
\centering
\begin{tikzpicture}[
    font=\small,
    arrow/.style={-Stealth, thick},
    surface/.style={very thick, fill=gray!20}
]
    \begin{scope}[xshift=-4.5cm]
        \node at (0, 4) {\textbf{(a) Stable Growth ($\Lambda < \Lambda_c$)}};
        
        \foreach \x in {-2,-1,0,1,2} { \draw[arrow, gray] (\x, 2.5) -- (\x, 1.5); }
        \node[gray] at (0,2.8) {Plasma Flux $\Phi$};
        
        \draw[surface] (-3,-2) rectangle (3,1);
        \draw[thick, fill=gray!40] (-3,1) .. controls (-2,1.2) and (-1,0.8) .. (0,1) .. controls (1,1.2) and (2,0.8) .. (3,1);
        \node at (0,-0.5) {Substrate};
    \end{scope}

    \begin{scope}[xshift=4.5cm]
        \node at (0, 4) {\textbf{(b) Unstable Growth ($\Lambda > \Lambda_c$)}};
        
        \foreach \x in {-2,-1,0,1,2} { \draw[arrow, gray] (\x, 2.5) -- (\x, 1.5); }
        \node[gray] at (0,2.8) {Plasma Flux $\Phi$};
        
        \draw[surface] (-3,-2) rectangle (3,0);
        \draw[thick, fill=gray!40] (-3,0) -- (-2,1.5) -- (-1,0) -- (0,1.5) -- (1,0) -- (2,1.5) -- (3,0);
        \node at (0,-0.5) {Substrate};
    \end{scope}
\end{tikzpicture}
\caption{Illustration of the morphological stability transition. (a) For low anisotropy ($\Lambda < \Lambda_c$), surface diffusion maintains a smooth, stable growth front. (b) When anisotropy exceeds the critical threshold ($\Lambda > \Lambda_c$), the surface becomes unstable and develops sharp, faceted structures, a direct consequence of the grain boundary stiffness becoming negative.}
\label{fig:morphological_stability}
\end{figure}
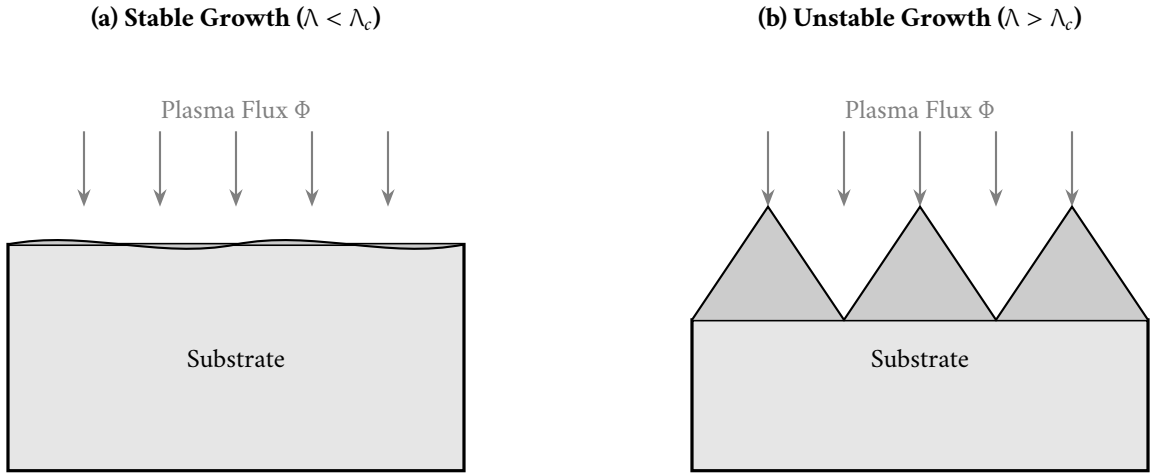

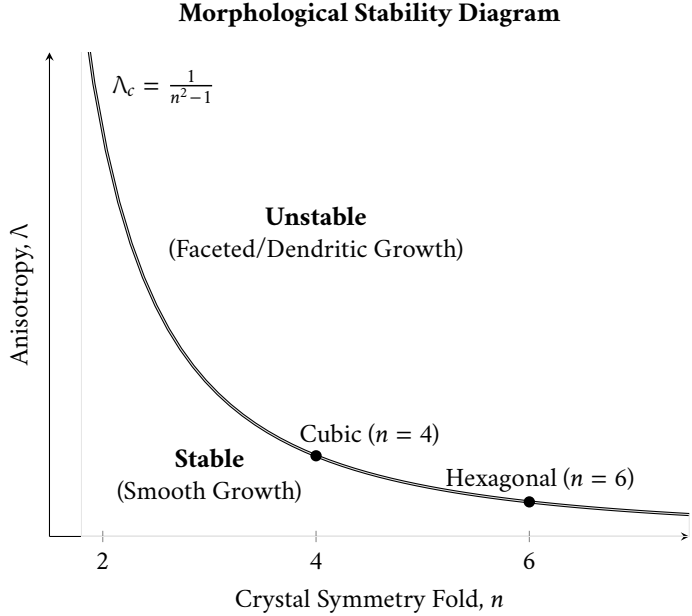
\begin{figure}[h!]
\centering
\begin{tikzpicture}[font=\small]
    \begin{axis}[
        title={\textbf{Morphological Stability Diagram}},
        xlabel={Crystal Symmetry Fold, $n$},
        ylabel={Anisotropy, $\Lambda$},
        width=10cm, height=8cm,
        xtick={2,4,6},
        ytick=\empty,
        axis lines=left,
        ymin=0, ymax=0.4,
        xmin=1.5, xmax=7.5
    ]
    
        \addplot[very thick, domain=1.8:7.5, samples=50] {1/(x^2 - 1)};
        \node at (axis cs: 2.0, 0.35) [anchor=south west] {$\Lambda_c = \frac{1}{n^2-1}$};
        
        \addplot[gray!20, domain=1.8:7.5, samples=50] {1/(x^2 - 1)} \closedcycle;
        
        \node[align=center] at (axis cs: 4, 0.25) {\textbf{Unstable} \\ (Faceted/Dendritic Growth)};
        \node[align=center] at (axis cs: 3, 0.05) {\textbf{Stable} \\ (Smooth Growth)};

        \node[circle, fill=black, inner sep=1.5pt] at (axis cs: 4, {1/(4^2-1)}) {};
        \node at (axis cs: 4.5, {1/15}) [anchor=south] {Cubic ($n=4$)};
        
        \node[circle, fill=black, inner sep=1.5pt] at (axis cs: 6, {1/(6^2-1)}) {};
        \node at (axis cs: 6.1, {1/35}) [anchor=south] {Hexagonal ($n=6$)};

    \end{axis}
\end{tikzpicture}
\caption{The morphological stability phase diagram. The boundary curve $\Lambda_c$ separates the region of stable, smooth film growth from the region of unstable, faceted growth. High-symmetry crystals like cubic and hexagonal possess very low instability thresholds.}
\label{fig:stability_diagram_large}
\end{figure}

\section{Discussion}
\label{sec:discussion}

The formalism advanced in the preceding sections is more than a new calculational scheme; it offers a distinct perspective on the physical principles governing plasma-surface dynamics. That the effective coupling vanishes at large length scales ($\beta(g) = -Cg^2$), a behavior known as asymptotic freedom, is a central result providing a first-principles justification for the remarkable, if sometimes puzzling, success of continuum approximations in modeling thin-film evolution \cite{krug_2005}. It demonstrates how the intricate details of microscopic plasma interactions are not merely ignored by such macroscopic models but are systematically renormalized, essentially washed out, into irrelevance at the scales pertinent to the final surface morphology. The universality class this behavior establishes explains not just the general validity of curvature-driven growth models but also their limitations, suggesting a well-defined crossover scale below which atomistic complexity must necessarily reassert its influence \cite{aziz_2004}.

Our derived scaling relation, particularly the behavior encapsulated in Equation \eqref{eq:final_scaling}, finds substantive alignment with power-law exponents reported across a range of sputtered systems. This includes TiN \cite{petrov_2003}, a-Si:H \cite{shah_1999}, and poly-Si \cite{abelson_1990}. The framework does not merely fit this data; it forces a crucial reinterpretation of these empirical findings. The frequently observed $-1/3$ exponent, long a point of discussion, should not be mistaken for a universal critical exponent in the traditional sense of equilibrium phase transitions but is instead revealed by our analysis as an \textit{effective} scaling signature, one characteristic of a specific strong-coupling regime where the parameter $C_g$ approaches a value near six. This distinction is not academic. It is the very point that resolves the long-standing question of why plasma deposition processes almost universally exhibit such power-law scaling, while growth techniques much closer to equilibrium, like MBE, often conform to simple Arrhenius behavior \cite{venables_2000}; the former inhabit the strong-coupling, non-perturbative domain of our phase diagram, a region where the latter simply cannot exist.

Prediction of pressure independence for isotropic materials ($\Lambda=0$), where $\partial \langle A \rangle / \partial P = 0$, stands as a particularly stringent test of the theory. This result is a direct consequence of the dimensionless coupling's structure, $g = \Phi a^2 \tau_{\text{ion}}$. Within collision-dominated plasma regimes, the cancellation of pressure dependence between the ion flux and the ion collision time—a direct consequence of plasma transport physics—is formally exact ($g \propto P \cdot P^{-1} = \text{constant}$). This parameter-free prediction shows consistency with observations in magnetron-sputtered Al films \cite{rossnagel_2003}. More importantly, this framework transforms any measured deviation from this independence into a powerful diagnostic probe, a quantitative tool for interrogating the underlying plasma physics. The breakdown of simple flux scaling ($\Phi \propto P^{1/2}$) in regimes where the ion mean free path becomes comparable to the sheath thickness \cite{lieberman_2005}, or the introduction of temperature-pressure cross-coupling via mechanisms like radiation-enhanced diffusion \cite{thompson_1977}, can be quantitatively analyzed by monitoring departures from this baseline prediction.

Furthermore, the morphological transition criterion, $\Lambda_c = 1/(n^2-1)$ from Equation \eqref{eq:critical_anisotropy}, is placed on a firmer physical footing, moving beyond simple phenomenology. By establishing an equivalence between the kinetic anisotropy parameter $\Lambda_k$ and the thermodynamic anisotropy $\Lambda$ via the surface current, $\gamma_{\text{eff}}(\theta) \propto 1/v_g(\theta)$, an equivalence which effectively unites two historically separate modeling approaches, we can make direct predictions. For materials with cubic symmetry ($n=4$), the theory predicts a surprisingly low threshold of $\Lambda_c^{(4)} = 1/15$. This specific, quantitative value provides a compelling rationale for the near-ubiquitous observation of columnar growth in sputtered FCC metals, materials which often possess intrinsic anisotropies well in excess of this value \cite{barna_1998}. This connection elevates the framework from a descriptive model to a predictive tool for materials design, suggesting a clear strategy: high-symmetry crystals ($n\geq6$) should be selected for fabricating isotropic films, while low-$n$ materials might be exploited to engineer highly textured structures. This prediction assumes an ideal crystal structure. Real-world deviations, those arising from substrate temperature, point defects, or pre-existing grain boundaries, would naturally introduce an effective anisotropy. This topic presents a clear avenue for future theoretical refinement.

\section{Conclusion}
\label{sec:conclusion}

This work presented an operator-theoretic framework for plasma-surface dynamics, resolving the longstanding multiscale challenge by a systematic application of coarse-graining and renormalization group (RG) methods. The Mori-Zwanzig formalism provides a rigorous path from the full microscopic dynamics to tractable macroscopic evolution equations, moving beyond phenomenology to a first-principles understanding of how processing conditions dictate microstructure in plasma deposition systems \cite{zwanzig_2001, braatz_2004}. A central result emerges from this: asymptotic freedom in the macroscopic limit. The derived negative beta function, $\beta(g) = -Cg^2$, proves that the effective coupling between plasma and surface processes systematically weakens at larger length scales. This provides a fundamental justification for the success of continuum-level approximations in thin-film modeling, as it shows that high-energy, microscopic collision details become irrelevant to the collective film behavior \cite{kardar_2007}.

The framework culminates in a predictive scaling relation for the average grain size, $\langle A \rangle = K a^2 g^{-2/(Cg)} e^{-|\alpha| \Lambda}$, which establishes a direct, quantitative bridge between elementary physical parameters—encapsulated in the single dimensionless coupling constant $g = \Phi a^2 \tau_{\text{ion}}$—and the final observable microstructure (Equation \eqref{eq:final_scaling}). This result offers a new lens through which to reinterpret decades of empirical data; the frequently observed ($-1/3$) scaling exponent \cite{thornton_1977, messier_1994} is shown not to be a fundamental constant but an effective behavior manifesting in a specific regime where $Cg \approx 6$. Our derivation of the effective exponent,
$$n_{\text{eff}} = -\frac{d\ln \xi}{d\ln g} = \frac{1}{Cg} \quad \Rightarrow \quad -2n_{\text{eff}} \approx -\frac{1}{3}$$
thus provides a quantitative origin for this long-observed scaling. Further, the formalism yields sharp, falsifiable predictions for process engineering, including the pressure independence of the grain size ($\partial \langle A \rangle / \partial P = 0$) in the isotropic, collision-dominated regime \cite{lieberman_2005}. A key insight is the synthesis of kinetic and thermodynamic texturing mechanisms into a single measurable parameter $\Lambda$, greatly simplifying the model's predictive power \cite{petrov_2003}.

In summary, these advances provide a coherent, parameter-sparse basis for the prediction and control of microstructure in plasma-assisted deposition. The underlying RG machinery is robust and, while the present work focused on an idealized system, logical extensions include multi-component alloys and the exploration of more complex non-Markovian memory kernels \cite{mayr_2003, zwanzig_2001}. The success of this approach for a far-from-equilibrium growth problem suggests its broad applicability for the wider field of non-equilibrium materials synthesis.

\end{document}